# A Model-data-driven Network Embedding Multidimensional Features for Tomographic SAR Imaging

Yu Ren
School of Information and Communication Engineering
University of Electronic Science and Technology of China
Chengdu, China
yuren1248@std.uestc.edu.cn

Xiaoling Zhang
School of Information and Communication Engineering
University of Electronic Science and Technology of China
Chengdu, China
xlzhang@uestc.edu.cn

Xu Zhan
School of Information and Communication Engineering
University of Electronic Science and Technology of China
Chengdu, China
zhanxu@std.uestc.edu.cn

Jun Shi
School of Information and Communication Engineering
University of Electronic Science and Technology of China
Chengdu, China
shijun@uestc.edu.cn

Shunjun Wei
School of Information and Communication Engineering
University of Electronic Science and Technology of China
Chengdu, China
weishunjun@uestc.edu.cn

Tianjiao Zeng
School of Information and Communication Engineering
University of Electronic Science and Technology of China
Chengdu, China
tzeng@uestc.edu.cn

*Abstract*—Deep learning (DL)-based tomographic SAR imaging algorithms are gradually being studied. Typically, they use an unfolding network to mimic the iterative calculation of the classical compressive sensing (CS)-based methods and process each range-azimuth unit individually. However, only one-dimensional features are effectively utilized in this way. The correlation between adjacent resolution units is ignored directly. To address that, we propose a new model-data-driven network to achieve tomoSAR imaging based on multi-dimensional features. Guided by the deep unfolding methodology, a two-dimensional deep unfolding imaging network is constructed. On the basis of it, we add two 2D processing modules, both convolutional encoder-decoder structures, to enhance multi-dimensional features of the imaging scene effectively. Meanwhile, to train the proposed multi-feature-based imaging network, we construct a tomoSAR simulation dataset consisting entirely of simulation data of buildings. Experiments verify the effectiveness of the model. Compared with the conventional CS-based FISTA method and DL-based $\gamma$-Net method, the result of our proposed method has better performance on completeness while having decent imaging accuracy.

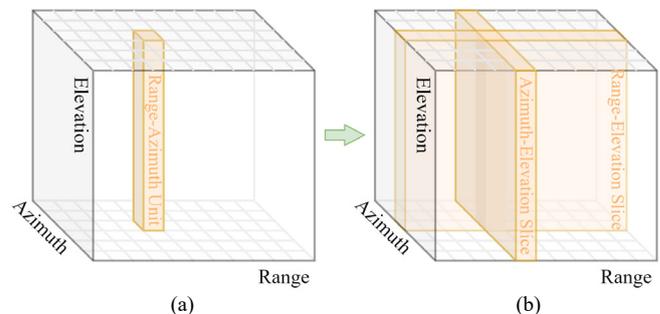

Fig. 1. Demonstrate of tomoSAR imaging using (a) deep unfolding methods and (b) multi-dimensional feature based methods.

## I. INTRODUCTION

Tomographic Synthetic aperture radar (TomoSAR) is a technology that realizes elevation inversion of imaging scenes through multi-view observation. Since the powerful ability to separate overlaid scatterers, it is widely used in fields such as earth remote sensing, especially in urban scenes [1].

Compressive sensing (CS)-based method is widely used in the tomoSAR imaging field. Conventional CS-based imaging algorithms have been extensively studied. By viewing tomo-SAR inversion as a sparse coding problem, such algorithms do tomoSAR by solving a linear inverse problem [2]. And since their super-resolution capabilities, better imaging performance can be obtained. However, conventional CS-based methods usually suffer from drawbacks like high computational cost, difficulty with manual parameter tuning, etc. [3].

Due to the powerful parameter learning ability of deep neural networks, the tomoSAR imaging algorithms based on deep learning (DL-based) have gradually begun to be studied. A typical class of algorithms uses a deep unfolding network to achieve elevation reconstruction, such as $\gamma$-Net [4]. It expands the iterations of conventional CS-based imaging algorithms into interconnected blocks and learns parameters. In a data-driven way, the parameters that were hard to fine-tune manually can now be learned from the training dataset, and the computational cost can be significantly reduced compared with the conventional CS-based method since the number of iterations can be dramatically reduced [5].

For the deep unfolding networks for tomoSAR imaging, they do one-dimensional reconstruction along the elevation direction for each range-azimuth unit, as shown in Fig.1 (a). Only 1D features are used in these methods. However, there is a specific correlation between different range-azimuth units in real imaging scenes. These multi-dimensional features have not been fully utilized in current DL-based methods. Meanwhile, CS-based methods usually use $L_1$ loss appropriately for prior constraints [2]. That results in only strong scattering points appearing in the imaging results. Characteristics of targets like the structure and continuity of imaging surface cannot be preserved.

To deal with these issues, we propose a tomoSAR imaging network to fully utilize multi-dimensional features. Firstly, we proposed a tomoSAR imaging model for multi-dimensional feature utilization like Fig.1 (b). Then, we implement it by integrating two convolutional encoder-decoder modules to a two dimensional deep-unfolding imaging network to process the range-elevation slices and azimuth-elevation slices, respectively, as shown in Fig. 2. After that, to complete the training of the proposed network, we constructed a simulation dataset entirely consisting of tomoSAR data of buildings. Finally, we train the proposed network on the simulation data to verify its performance. Experiments demonstrate the effectiveness of the method. Compared with $\gamma$-Net and conventional FISTA, the results of the proposed method have better continuity.

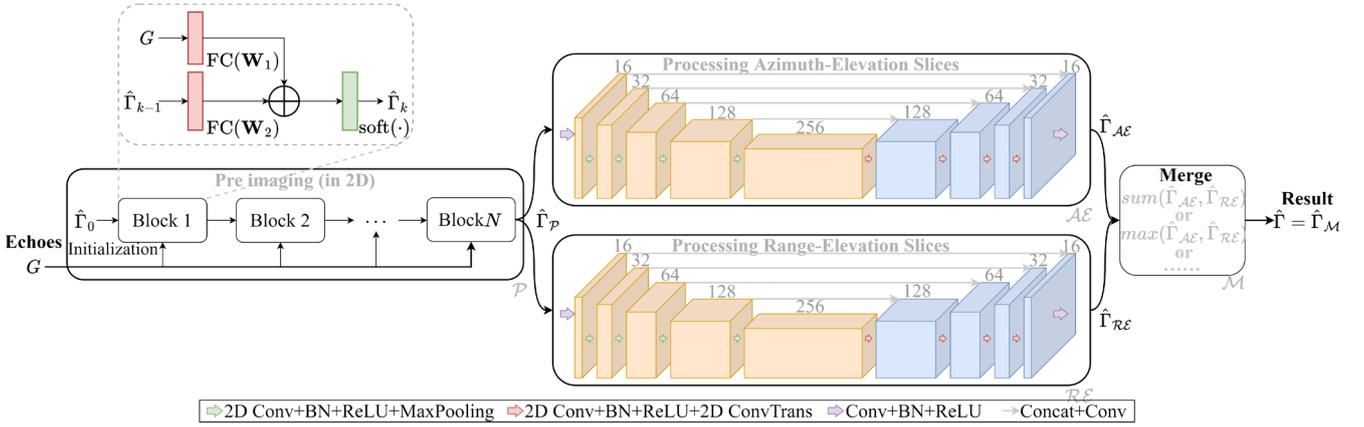

Fig. 2. Architecture the proposed tomoSAR imaging network.

## II. METHODOLOGY

### A. TomoSAR Imaging with Unfolding Network

The acquisition of echoes $g$ by tomoSAR is a process of linear observation of the imaging scene [2], which can be discretely expressed as

$$g = A\gamma + n \quad (1)$$

Where $g$ is the echoes received from the tomoSAR system, $\gamma$ represents the scattering characteristics, $A$ is the measurement matrix which is determined by the tomoSAR system and the imaging scene, and $n$ is noise. For tomoSAR, due to limited baseline length, the imaging is usually to invert high-dimensional $\gamma$ from low-dimensional $g$.

Conventionally, CS-based algorithms do imaging in an iterative manner for each range-azimuth unit. For example, the ISTA algorithm [6] does that by computing the following equation iteratively,

$$\hat{\gamma}_k = h_\theta \left( \hat{\gamma}_{k-1} - \mu A^H (A\hat{\gamma}_{k-1} - g) \right) \quad (2)$$

where, $h_\theta(\cdot) = \text{sign}(\cdot) \max(|\cdot| - \theta, 0)$ is a soft thresholding function with a $\theta$ threshold, $\mu$ is the iteration step size, and $\hat{\gamma}_k$ and $\hat{\gamma}_{k-1}$ are the reconstruction results of $k$-th and $(k-1)$-th iteration.

However, several hyper-parameters are usually needed to set manually in such CS-based algorithms. In practical applications, it usually takes a long time to do parameter tuning to get a satisfying result. Furthermore, it needs many iterations to get a convergent result. That usually costs much time to do that calculation.

Recently, $\gamma$-Net is proposed to achieve tomoSAR inversion using deep learning. By rewriting the iteration equation to

$$\hat{\gamma}_k = h_\theta(W_1 g + W_2 \hat{\gamma}_{k-1}) \quad (3)$$

$\gamma$-Net converts the above ISTA iteration to the unfolding network and learns parameters $\theta, W_1$ and $W_2$, or simpler, inputs a fixed $A$ and learns parameters $\mu, \theta$. In a data-driven way, corresponding parameters can be learned from simulation data instead of being set manually. And this method can greatly reduce the number of iterations [5], which is also the number of blocks in $\gamma$-Net, to achieve a significant reduction in computational consumption.

### B. Multi-dimensional Features based Imaging Network

TomoSAR imaging using an unfolding network still performs 1D inversion operations like conventional CS-based methods, i.e., it does tomoSAR reconstruction for each range-azimuth unit individually without considering the association between adjacent resolution units. It usually results in that only strong scattering points being found in the result of $\gamma$-Net. That causes some phenomena like weak target miss, target continuity loss, etc.

By incorporating multi-dimensional features in the 3D imaging scene, imaging calculations can be done based on the features of the entire imaging scene instead of the original 1D features.

To fully utilize multi-dimensional features in 3D imaging scenes, we need to construct a 3D operator $\mathcal{O}_{3D}$ to do tomo-SAR imaging. Intuitively, we can use a 3D feature extraction model directly, like a 3D-convolutional network, to do that. Nevertheless, unfortunately, since 3D operations are often computationally expensive, it has high requirements on the machine needed for training.

According to our observations of simulated and measured data, we found that the height of imaging targets does not change drastically alone range direction or azimuth direction. Therefore, we can perform two 2D processing along the range direction and the azimuth direction, respectively, to fully utilize 3D features.

Thus, a multi-dimensional-based network is proposed, as shown in Fig. 2. It consists of three parts.

- The first part performs pre-imaging. By splicing echoes in the same range/azimuth unit, we can get 2D echo slices along the azimuth-elevation/range-elevation direction. Then the pre-imaging part can perform 2D pre-imaging for every 2D slice.
- The second part performs 2D feature processing along

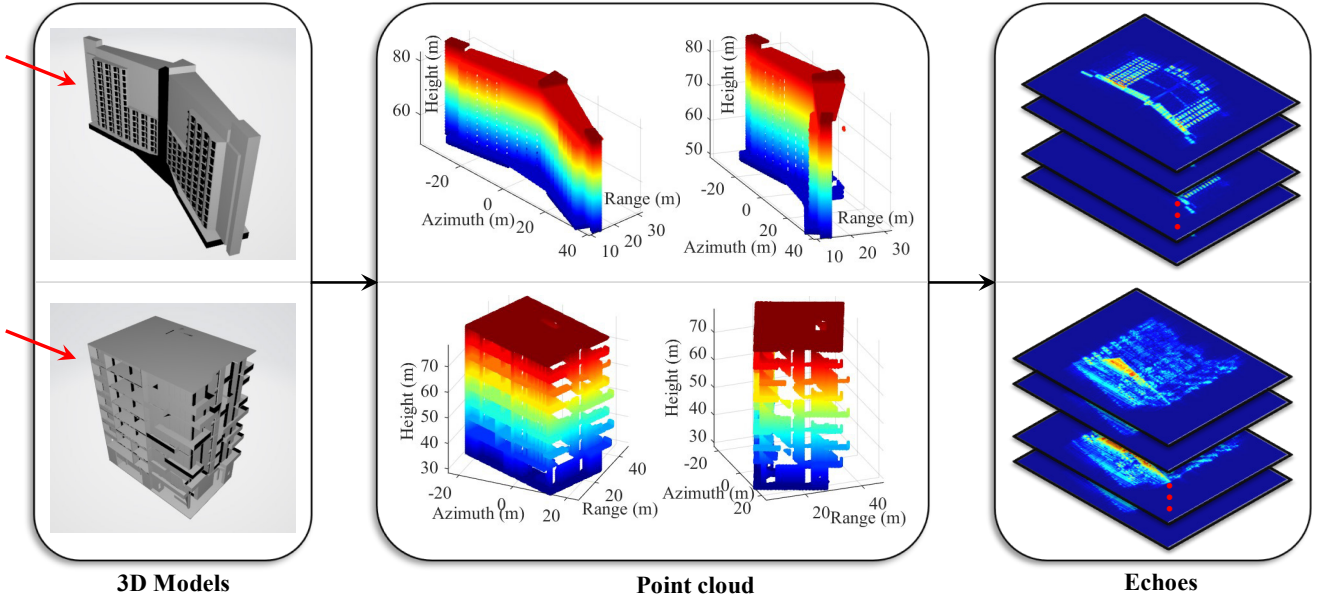

Fig. 3. Demonstration of simulation steps and examples of simulation tomoSAR dataset.

two directions, which are the range-elevation and azimuth-elevation directions, to make full use of multi-dimensional features. It uses 2D features to process each pre-imaged imaging result slice in a 2D manner to achieve effects like fault surface filling, outlier removal, etc.

- The final part merges the results processed by the previous 2D parts. All the operations are based on 2D slices.

For each part of the proposed network, the processes can be formulated as follows.

The pre-imaging part is done by $\mathcal{P}$ operator. By slicing the echo data along the range/azimuth direction, we can get echo slices $G$. By inputting echoes slices, $\mathcal{P}$ do pre-imaging in an iterative way, like unfolding network, which can be formulated as

$$\hat{\Gamma}_{\mathcal{P}} = \mathcal{P}(G) \qquad (4)$$

For each block in $\mathcal{P}$, it does the following calculation.

$$\hat{\Gamma}_k = h_\theta \left( \hat{\Gamma}_{k-1} - \mu \mathbf{A}^H \left( \mathbf{A} \hat{\Gamma}_{k-1} - G \right) \right) \qquad (5)$$

where $\hat{\Gamma}_k$ is the result of $k$-th iteration, i.e., the output of block $k$ in the unfolding pre-imaging network, $G$ is the echoes slice alone azimuth-elevation direction. Compared with the step in (2), it should be noted that the iteration flow here is 2D.

The second part performs 2D processing along the range-elevation direction and azimuth-elevation direction. They process the pre-imaged results $\hat{\Gamma}_\mathcal{P}$ from the 2D perspective in both range and azimuth directions. We can define them by 2D operators $\mathcal{AE}$ and $\mathcal{RE}$ and the result of them by $\hat{\Gamma}_{\mathcal{AE}}$ and $\hat{\Gamma}_{\mathcal{RE}}$.

After that, $\Gamma_{\mathcal{AE}}$ and $\Gamma_{\mathcal{RE}}$ can be merged by the final part using $max(\Gamma_{\mathcal{AE}}, \Gamma_{\mathcal{RE}})$, $sum(\Gamma_{\mathcal{AE}}, \Gamma_{\mathcal{RE}})$ or other operations. Denote the merging method as $\mathcal{M}$, then the entire imaging operator we proposed can be expressed as

$$\hat{\Gamma} = \mathcal{O}(G) = \mathcal{M}\left( \mathcal{AE}(\mathcal{P}(G)) + \mathcal{RE}(\mathcal{P}(G)) \right) \qquad (6)$$

where,

- $\mathcal{AE}$ and $\mathcal{RE}$ are the operators to process azimuth-elevation slices and range-elevation slices, respectively, to take full use of the multi-dimensional features along those two directions. As the architecture of the encoder-decoder has shown superior feature extraction and mapping ability in multiple areas [7]–[9], we take this architecture with skip connections in the method, as shown in the right part of Fig. 2. Notably, the two branches have different network parameters to learn.

- $\mathcal{M}$ means to merge the results after processing by $\mathcal{AE}$ and $\mathcal{RE}$.

### C. Simulation Dataset

Training data used in most of the existing DL-based tomoSAR imaging networks are simulated data of point targets [3], [4]. There are a few drawbacks to it: (1) When simulating tomoSAR data of point targets, the distribution of the targets along elevation direction is manually set. It does not necessarily conform to the distribution in the actual scene. (2) When simulating data in this way, each range-azimuth unit is considered individually. The association between adjacent units cannot be utilized from the multi-dimensional perspective. (3) Multiple scattering effect cannot be considered in this simulation mode.

Therefore, we construct a tomoSAR simulation dataset consisting entirely of building CAD models. The simulation procedure is referred to the simulation method of RaySAR [10], a SAR image simulation software open-sourced by DLR. The specific steps are as follows:

- Select several 3D models of buildings. For the richness of the data, we have selected various types of buildings, including changes in height, shape, etc.

- Generate a point cloud for each building from the perspective of the SAR sensor. In order to make the simulated data more realistic, we consider the occlusion relationship when generating the point cloud. The middle part in Fig. 3 demonstrates our occlusion considerations.

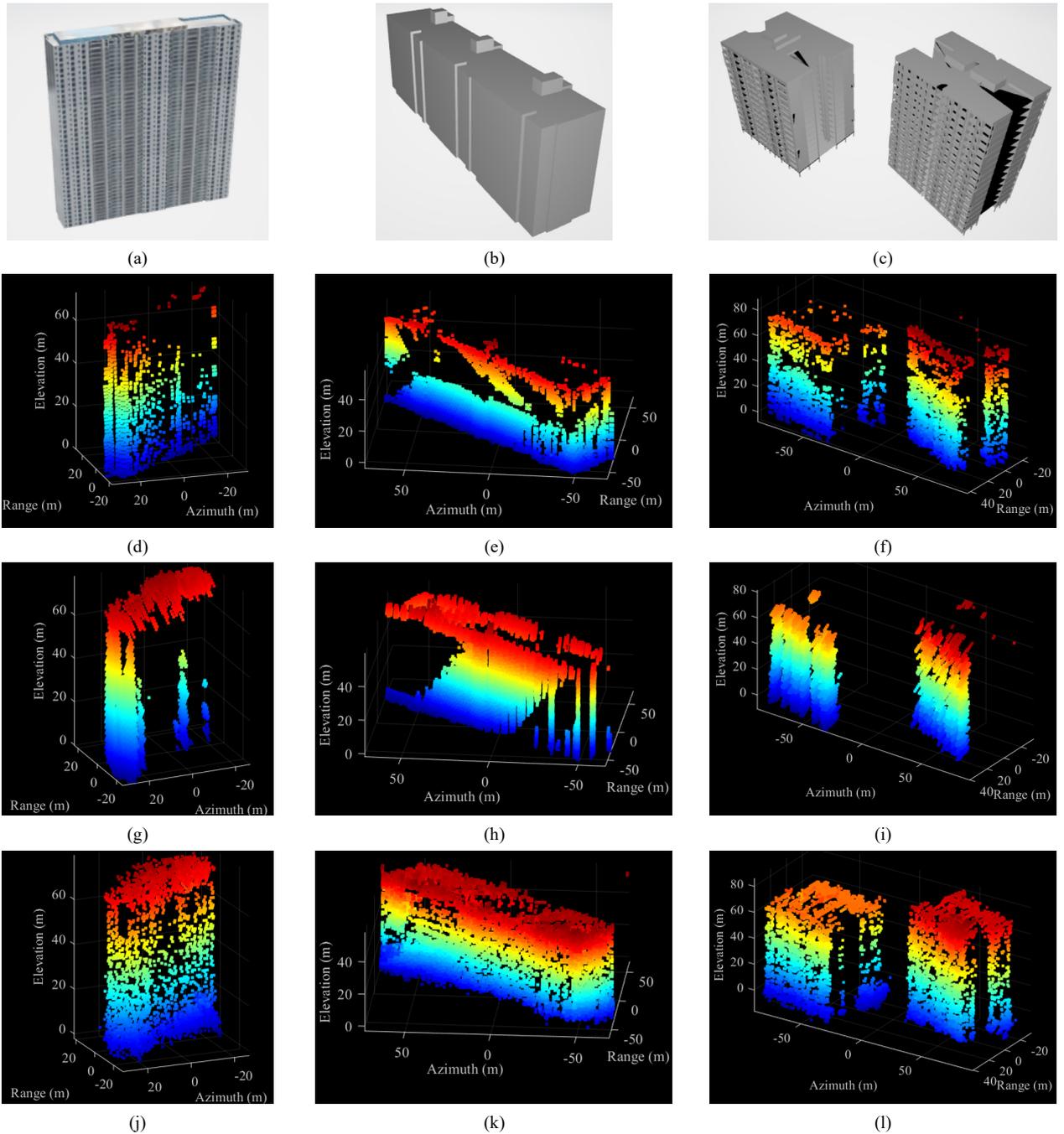

Fig. 4. Imaging results on simulated data of (a – c) buildings using (d – f) FISTA, (g – i) $\gamma$-Net and (j – l) proposed method.

- Generate the echoes from each flyover view and ground truth in RaySAR way.

A demonstration of simulation steps and several examples related to the simulation dataset is shown in Fig. 3.

## III. EXPERIMENT

### A. Dataset

We have made such simulation data for 54 models of the building. For the simulation of each model, there are 152, 200, and 128 resolution units in range, azimuth, and elevation direction, respectively. 11 flyovers are used in our simulation with approximately uniform distribution in the range of 0 m to 1.9896 m. We set the height of the reference baseline to 1736.9668 m and set the center range distance $R_0$ to 2040.3406 m. The radar incidence angle of the reference baseline is set to $31.6453°$. And the simulated tomoSAR system works at X-band.

### B. Network Implementation

For the pre-imaging part, we implemented it using complex fully connected (FC) layers. Each block consists of two FC layers and one soft operation as an activation function. In our implementation, the pre-imaging part is composed of 5 blocks.

For the second part of the proposed network, both $\mathcal{AE}$ and $\mathcal{RE}$ parts are implemented by convolutional encoder-decoder with the same structure. The encoding and decoding parts are composed of 5 stages. Each stage do 2D conv+2D batch normalization+ReLU twice and max-pooling/2D-transpose-convolution for encoder/decoder respectively once. And at the end of $\mathcal{AE}$ and $\mathcal{PE}$, they both use 2D convolution layers to output the result. The detailed structure of each stage is shown in Fig. 2.

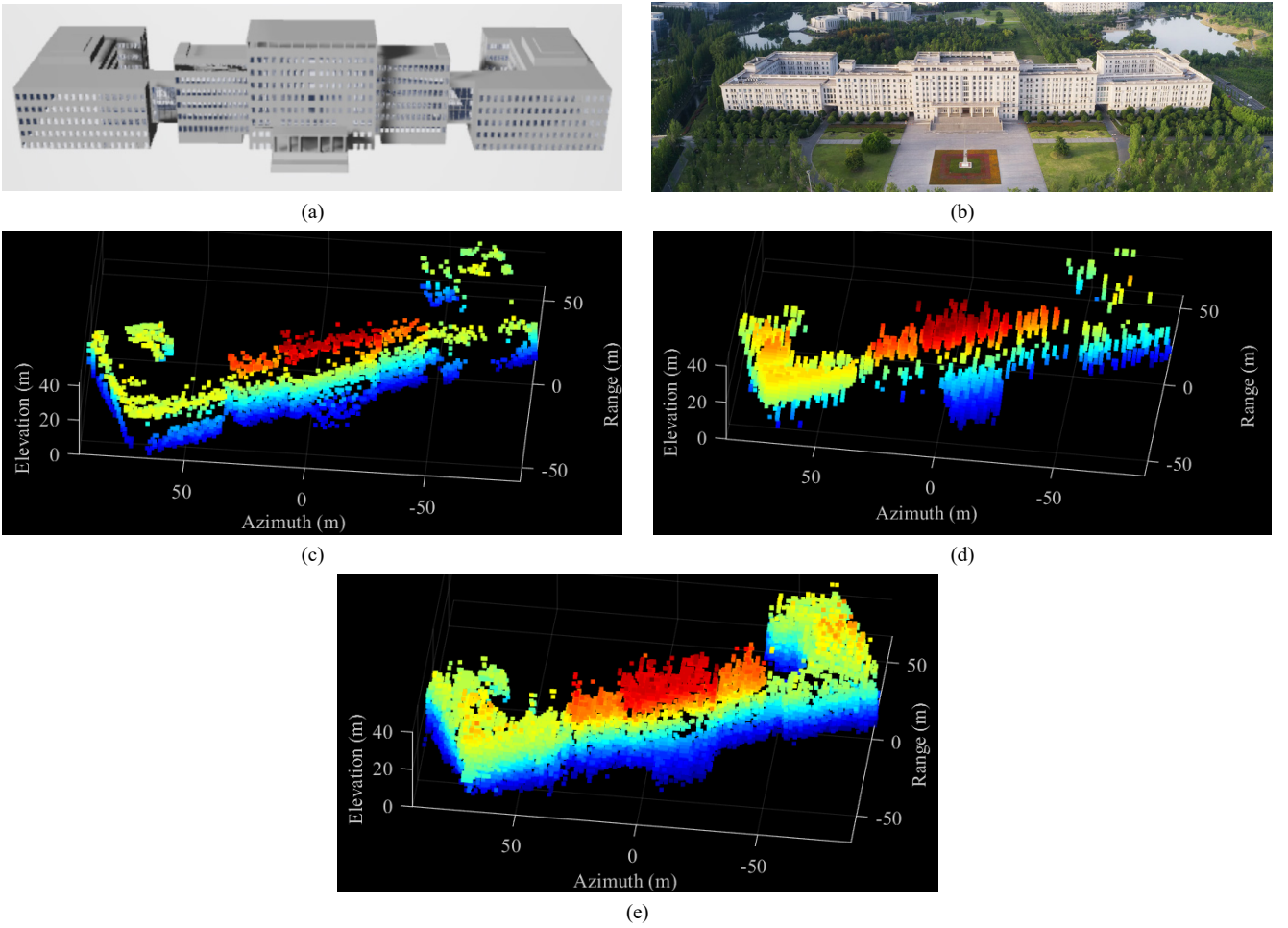

Fig. 5. Simulation experiment of the administration building of University of Electronic Science and Technology of China, UESTC. (a) The 3D model of the building. (b) The optical picture of the building. (c~e) Imaging results demonstration using (c) FISTA, (d) $\gamma$-Net and (e) proposed method.

And, for the realization of part $\mathcal{M}$, we choose $max(\hat{\Gamma}_{\mathcal{AE}}, \hat{\Gamma}_{\mathcal{RE}})$ to achieve the final imaging result.

### C. Model Training

The training of the proposed model requires two stages. Firstly, the pre-imaging part needs to be pre-trained. This allows parameters such as the measurement matrix of the pre-imaging model to be as accurate as possible. In the second stage, the full model starts to be trained. And in this process, the parameters of the pre-imaging part are properly fine-tuned based on the 2D processing modules.

In our training, during the pre-training stage, a total of 30 epochs were used to bring the pre-imaging part to convergence. The batch size was set to 128, and the learning rate was $l_{r1} = 10^{-5}$, and we used Adam optimizer.

In the second stage of the entire training procedure, we use a total of 50 epochs. The batch size was set to 32, and the learning rate was $l_{r2} = 10^{-4}$, and we also used Adam optimizer.

### D. Loss Function Design

In order to ensure the accuracy of the result after $\mathcal{P}$ and retain as many features as possible, We only use $L_2$ loss function to ensure the accuracy of the result of the pre-imaging part, which is

$$\mathcal{L}_{pre} = \|\hat{\Gamma}_{\mathcal{P}} - \Gamma^*\|_2^2 \quad (7)$$

For the second stage of model training, we use both $L_2$ and $L_1$ loss functions to constrain the accuracy and sparsity of the reconstruction results, that is

$$\mathcal{L} = \|\hat{\Gamma} - \Gamma^*\|_2^2 + \lambda \cdot \|\hat{\Gamma}\|_1 \quad (8)$$

Since the two-dimensional processing module can handle noise and other problems well, we only add weak $L1$ constraints, so in our experiment, we set $\lambda = 0.01$.

## IV. RESULT

Several examples for visual comparison, as shown in Fig. 4 and Fig. 5, demonstrate imaging results of FISTA, $\gamma$-Net, and the proposed method. We can clearly see that the results of the proposed method have better continuity and are able to maintain a more complete structure of the imaging buildings.

We also conduct comparative experiments on the three scenes in Fig.4, using FISTA, $\gamma$-Net, and the proposed method, respectively, to evaluate the performance of the proposed method through completeness and accuracy metrics [11]. The result is shown in TABLE Ⅰ.

It can be seen from the results in TABLE Ⅰ that the completeness of the imaging results of the proposed method can achieve the best among all of the tested three models. The proposed method has the best performance in preserving the continuity and structure of imaging results. Moreover, among the tested models, the imaging results of the proposed method are also optimal in terms of accuracy.

TABLE I. Comparison experiment for 3 models in Fig. 4 with completeness and accuracy.

| Model | Method | Completeness | Accuracy |
|---|---|---|---|
| 1 | FISTA | 2.1312 | 2.2303 |
|   | $\gamma$-Net | 3.0741 | 1.6066 |
|   | Proposed | **0.8584** | **1.3283** |
| 2 | FISTA | 1.8891 | 1.6464 |
|   | $\gamma$-Net | 4.9143 | 1.4405 |
|   | Proposed | **1.0933** | **1.1863** |
| 3 | FISTA | 5.5364 | 1.7959 |
|   | $\gamma$-Net | 3.4290 | 1.9901 |
|   | Proposed | **0.9238** | **1.3260** |

## V. CONCLUSION

In this work, faced with the problem of insufficient utilization of multi-dimensional features in the imaging process, a two-dimensional deep unfolding imaging network is constructed. On the basis of it, we consider adding two 2D processing modules to perform azimuth-height slices and distance-height slices, respectively, to add multi-dimensional features into the tomoSAR imaging procedure. Meanwhile, we construct a dataset consisting entirely of simulation data of buildings to do the network training. Experiments on simulation data demonstrate the effectiveness of the proposed method. Compared with conventional FISTA and deep unfolding imaging network $\gamma$-Net, our proposed method performs better in preserving the continuity of imaging targets.